\documentclass{ieeeaccess}
\usepackage{cite}
\usepackage{algorithmic}
\usepackage{graphicx}
\usepackage{textcomp}
\usepackage{multirow}
\usepackage{comment}
\usepackage{cite}
\usepackage{url}
\usepackage{amsmath,amsthm,amssymb,amsfonts}
\usepackage{mathtools}
\usepackage{inputenc}
\usepackage{float}
\usepackage{physics}
\usepackage{enumitem}

\usepackage{algorithm} 
\newcommand{\bd}[1]{\mathbf{{#1}}}\textit{}
\newcommand{\paren}[1]{\left({#1}\right)}

\newcommand{\bracket}[1]{{\left [{#1}\right ]}}
\newtheorem{theorem}{Theorem}
\newtheorem{lemma}{Lemma}
\newtheorem{corollary}{Corollary}
\DeclareMathOperator{\E}{\mathbb{E}}
\usepackage{caption}
\newcommand{\ist}[1]    {{#1}^{\underline{ \text{st}}}}
\newcommand{\ith}[1]    {{#1}^{\underline{ \text{th}}}}
\def\BibTeX{{\rm B\kern-.05em{\sc i\kern-.025em b}\kern-.08em
    T\kern-.1667em\lower.7ex\hbox{E}\kern-.125emX}}
\begin{document}
\history{Date of publication xxxx 00, 0000, date of current version xxxx 00, 0000.}
\doi{10.1109/ACCESS.2023.DOI}

\title{Low Complexity Iterative 2D DOA Estimation in MIMO Systems}
\author{\uppercase{Md Imrul Hasan}\authorrefmark{1}, \IEEEmembership{Student Member, IEEE},
\uppercase{Mohammad Saquib}\authorrefmark{1}
\IEEEmembership{Senior Member, IEEE}}
\address[1]{Department of Electrical Engineering, The University of Texas at Dallas, Richardson, 
TX 75080 USA}
\tfootnote{}


\markboth
{Author \headeretal:  Low Complexity Iterative 2D DOA Estimation in MIMO Systems }
{Author \headeretal: Low Complexity Iterative 2D DOA Estimation in MIMO Systems}
\corresp{Corresponding author: Md Imrul Hasan (mdimrul.hasan@utdallas.edu) and Mohammad Saquib (saquib@utdallas.edu)}

\begin{abstract}
Multiple-input multiple-output (MIMO) systems play an essential role in direction-of-arrival (DOA) estimation. A large number of antennas used in a MIMO system imposes a huge complexity burden on the popular DOA estimation algorithms, such as MUSIC and ESPRIT due to the implementation of eigenvalue decomposition. This renders those algorithms impractical in applications requiring quick DOA estimation. Consequently, we theoretically derive several useful noise subspace vectors when the number of signal sources is less than the number of elements in both the transmitter and receiver sides. Those noise subspace vectors are then utilized to formulate a 2D-constrained minimization problem, solved iteratively to obtain the DOAs of all the sources in a scene. The convergence of the proposed iterative algorithm has been mathematically as well as numerically demonstrated. Depending on the number of iterations, our algorithm can provide significant complexity gain over the existing high-resolution 2D DOA estimation algorithms in MIMO systems, such as MUSIC, while exhibiting comparable performance for a moderate to high signal-to-noise ratio (SNR).

\end{abstract}


\titlepgskip=-15pt

\maketitle

\section{Introduction}
\label{sec:introduction}

\PARstart{D}{irection} of arrival (DOA) is an important concept in a wide range of areas, such as wireless communications, sonar, radar, geophysics, defense operations, seismology, etc \cite{a01, ruan2022doa, hasan2022robust, hasan2022low, mp2, zimerman2020colored}. In these applications, DOA estimation is used, 1) to guide the beam in order to improve signal quality, suppress interference, and reduce the power for transmission, 2) to localize a target, which is critical for navigation, communication, surveillance, and tracking, 3) to detect the origin of seismic waves, which allows the exploration of the Earth's interior, predicting earthquakes, and so on. 

Multiple Input Multiple Output (MIMO) technologies play an essential role in DOA estimation by allowing diversity, a high spatial multiplexing gain, and improved spatial resolution of array elements. Since a MIMO system utilizes many antennas collected in one panel and permits numerous paths for the signal to reach the receiver, it helps to reduce the impact of fading and interference. Another advantage of a MIMO system is the formulation of the virtual array \cite{di20202,lee2021virtual,chen2013novel}, which describes the practical spatial locations from which the target reflectivity gets sampled. That means if the monostatic transmitter and receiver elements operate independently at each virtual array element location, those will provide the exact phase measurements as the physical configuration. The virtual array in a MIMO system enhances the original aperture size, which as a result, increases the degrees of freedom (DOF)  \cite{bliss2003multiple}. These attributes offer high accuracy and robustness in the DOA estimation with a MIMO system compared to a common phased array system.

 In the literature, extensive studies have been performed, and numerous algorithms have been developed to estimate the DOAs, i.e. multiple signal classification (MUSIC) \cite{b1}, maximum-likelihood (ML) \cite{b2}, Capon \cite{b01}, estimation of signal parameters via rotational invariance techniques (ESPRIT) \cite{b4}, Min-Norm \cite{b5}, etc. Among all these methods, traditional subspace-based estimation algorithms, such as MUSIC, and ESPRIT are especially noteworthy due to their high precision and resolution in DOA estimation\cite{santiago2013noise, stoeckle2015doa}. This noise subspace-based estimation (NISE) algorithm  \cite{santiago2013noise} iteratively solves for each source's elevation angle, potentially yielding (depending on the number of iterations) lower complexity than existing DoA estimation algorithms, such as Fast Root-MUSIC (FRM) \cite{FRM1997}.

 MIMO systems employ 2D antenna elements due to their 3D beamforming benefits. Therefore, the 2D DOA estimation in a MIMO system is of great interest. For the past few years, a number of 2D DOA estimation techniques have been introduced based on MUSIC and ESPRIT algorithms. For example, a MUSIC-based algorithm is proposed in \cite{yang2014low} which estimates the DOAs in two stages, but still requires eigenvalue decomposition and peak searching to estimate the angles. A DOA estimation algorithm based on 2D MUSIC in MIMO systems is proposed in \cite{meng2016low}, however, it uses spatial and eigenvalue decomposition. A 2D unitary ESPRIT algorithm is introduced in \cite{jian2006two}, but it fails to achieve a high precision because of the constraint concerning the array aperture. Moreover, the eigenvalue decomposition (EVD) of the covariance matrix needs to be computed. This leads to a high computational complexity while using large-scale antenna arrays since the array in MIMO systems can contain hundreds of elements\cite{stoeckle2015doa}. Moreover, as the electromagnetic environment is getting increasingly complex every day, the computational burden of these algorithms in MIMO systems poses an issue of great concern, especially in a real-time environment where a quick estimation of the DOAs is required \cite{suk2018low, xiao2016new, wang2012low,liu2021low,feng2018target}.  The low complexity NISE algorithm \cite{santiago2013noise} is capable of quickly estimating 1D DOA employing a uniform linear array (ULA). Unfortunately, it is not suitable for a MIMO system. 
 
In light of the above, we seek a low-complexity quick 2D DOA estimation technique. For a MIMO system, closed-form expressions of several non-signal subspace eigenvectors are derived using the algebraic expressions of the non-signal subspace developed in \cite{santiago2013noise} for a ULA. Those eigenvectors contain information on all the DOAs in a scene and are utilized to formulate a 2D-constrained minimization problem. We solve this problem using an iterative 2D DOA estimation algorithm, namely iDEA. Our analysis shows that in a noise-free scenario, as the number of iterations approaches infinity iDEA's estimates approach the true values of the DOAs. Since iDEA is an iterative algorithm, it allows performing a trade-off between the performance and complexity by controlling the number of iterations. An extensive numerical study is carried out to investigate the performance of iDEA against the 2D MUSIC algorithm. This study also includes a thorough complexity comparison between those two.

The rest of the paper is organized as follows: Section II discusses the System Model for MIMO systems. In Section III, eigenvectors of the non-signal subspace for MIMO systems are formulated and later demonstrated how these vectors could be utilized to determine the  $\ith{K}$ source, given we have prior information of the other $K-1$ sources in the scene. Section IV describes the iDEA algorithm for MIMO systems to estimate all the $K$ sources sequentially while using an arbitrary number of antennas. In Section V, the convergence proof of the algorithm is presented, and a thorough performance analysis of iDEA w.r.t. the 2D MUSIC algorithm is performed for a MIMO system in Section VI. Finally, Section VII contains the concluding remarks of this paper.

\textit{Notations:} We use lowercase and uppercase bold letters to denote vectors and matrices, respectively. Lowercase letters in italics are used to represent scalars. The notation * refers to the complex conjugate of a scalar, $[\cdot]^T$ refers to transpose, and $[\cdot]^\dag$ denotes the Hermitian of a matrix. In this paper, $\otimes$ is used to portray the Kronecker product.
  \begin{figure}[t!]
 \centering
 \includegraphics[scale=0.90]{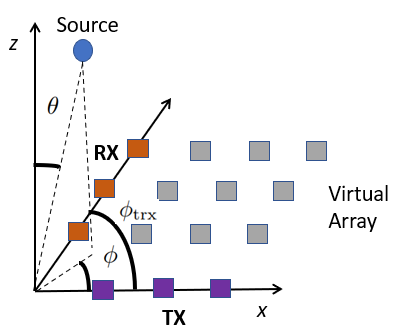}
 \caption{ 2D DOA estimation in MIMO system.}
    \label{URA}
\end{figure} \\

\textit{Properties of Kronecker Products (PKP) :} In this paper, the following properties of the Kronecker Products\cite{schacke2004kronecker} are used to formulate and prove the convergence of the proposed algorithm, iDEA.
\begin{itemize}
      \item PKP 1: $\paren{\mathbf{A} \otimes \mathbf{B}}^\dag =  \mathbf{A}^\dag \otimes \mathbf{B}^\dag$ 
     \item PKP 2:  $k\bd{A}\otimes \bd{B}$ = $\bd{A}\otimes k\bd{B} = k\paren{\bd{A}\otimes \bd{B}}$
   
     \item PKP 3:
     \begin{itemize}
     \item $\bd{A}\otimes \paren{\bd{B}+\bd{C}} = \bd{A}\otimes \bd{B}+\bd{A}\otimes \bd{C}$ 
     \item $\paren{\bd{A}+\bd{B}} \otimes \bd{C} = \bd{A}\otimes\bd{C}+ \bd{B}\otimes\bd{C}$
     \end{itemize}
    
    \item PKP 4: $\paren{\mathbf{A} \otimes \mathbf{B})(\mathbf{C} \otimes \mathbf{D}}$ = $\mathbf{(AC)} \otimes \mathbf{(BD)}$
\end{itemize}

\section{System Model} Let us consider a MIMO system, where $N_\mathrm{tx}$ transmitters are placed linearly along the $\mathrm{x}-$ axis, and $N_\mathrm{rx}$ receivers are positioned linearly at an angle of $\phi_{\mathrm{trx}}$ with the transmitters; see Fig. \ref{URA}. At
the transmit site, $N_\mathrm{tx}$ different narrowband waveforms are emitted simultaneously, which have identical bandwidth and center frequency but are temporally orthogonal. In each receiver, the signals are processed for all of the transmitted waveforms. Assume the number of noncoherent far-field targets in the scene, $K$ is previously known \cite{zhen2010method}, and these signals impinge upon the MIMO with wavelength $\lambda$, where $K<N_\mathrm{tx}$ and $ N_\mathrm{rx}$. 
The azimuthal angle of the $k^{\underline{ \text{th}}}$ user, $0^{\circ}\leq \phi_k< 360^{\circ}$ is measured counterclockwise from the $\mathrm{x}-$ axis, and its elevation angle, $0^{\circ}\leq\theta_k\leq 90^{\circ}$ is measured downward from the $\mathrm{z}-$ axis. Let us denote the steering matrix for the transmitters, and the receivers as $\mathbf{A}_\mathrm{tx}$, and $\mathbf{A}_\mathrm{rx}$, respectively. Therefore, for $K$ number of targets, we can write $$\mathbf{A}_\mathrm{tx}=\bracket{\mathbf{a}_{\mathrm{tx},1}, \ldots, \mathbf{a}_{\mathrm{tx},K}},$$  and 
$$\mathbf{A}_\mathrm{rx}=\bracket{\mathbf{a}_{\mathrm{rx},1}, \ldots, \mathbf{a}_{\mathrm{rx},K}}.$$ 
Given the geometry, and the uniform array element spacing $d \leq \lambda/2$, the ${i}^{\underline{ \text{th}}}$ element of the steering vector $\mathbf{a}_{\mathrm{tx},k}$,
\begin{equation}
    a_{\mathrm{tx},i,k}=\exp\{j\times i\epsilon_{\mathrm{tx},k}\}; \, i=1,\ldots,N_\mathrm{tx} \, \, k=1,\ldots,K,
\end{equation}
and  the ${l}^{\underline{ \text{th}}}$ element of the steering vector $\mathbf{a}_{\mathrm{rx},k}$,
\begin{equation}
    a_{\mathrm{rx},l,k}=\exp\{j\times l\epsilon_{\mathrm{rx},k}\}; \, l=1,\ldots,N_\mathrm{rx} \, \, k=1,\ldots,K,
\end{equation} where 
$$\epsilon_{\mathrm{tx},k}=- \paren{2\pi d/\lambda}\sin \theta_k \cos \phi_k,$$ 
$$\epsilon_{\mathrm{rx},k}=-\paren{2\pi d/\lambda} \sin \theta_k \cos (\phi_k-\phi_{\mathrm{trx}}).$$

Now, the virtual steering vector for the MIMO is given by the following relation \cite{di20202}:
\begin{equation}
    \mathbf{A}=\bracket{\mathbf{A}_{1},\ldots,\mathbf{A}_{K}}\,,
    \label{str}
\end{equation}
where $$\mathbf{A}_{k} = \mathbf{a}_{\mathrm{tx},k} \otimes \mathbf{a}_{\mathrm{rx},k}$$
is the contribution from the $\ith{k}$ user. Therefore, at the $m^{\underline{\mathrm{th}}}$ snapshot, the outputs of all the matched filters in all the receivers can be expressed as \cite{duofang2008angle}
\begin{equation}
    \mathbf{x}(m)=\mathbf{A}\mathbf{s}(m)+\mathbf{w}(m),
    \label{rs}
\end{equation}
where $\mathbf{x}(m) \in C^{N \times 1}$, $\mathbf{A} \in C^{N \times K}$, $\mathbf{s}(m) \in C^{K \times 1}$, and $\mathbf{w}(m) \in C^{N \times 1}$. Here, $\mathbf{s}(m)$ is a column vector consisting of the phases and amplitudes of the $K$ sources, vector $\mathbf{w}$(m) contains complex additive white Gaussian noise (AWGN) samples with average power $\sigma^2/2$ per dimension and spatially and temporally independent of $\mathbf{s}(m)$, and by definition, $N=N_\mathrm{tx} N_\mathrm{rx}$. In the next section, we will develop the model for the eigenvectors in the non-signal subspace of the auto-correlation matrix $\mathbf{R}$ based on the received signal vector (\ref{rs}). 

\section{Non-Signal Subspace Analysis}
The auto-correlation matrix, 
 \begin{align}
 \mathbf{R}& =\E
 \{\mathbf{{x}}(m)\mathbf{{x}}(m)^\dag\}\nonumber\\
 &=\mathbf{A}\E \{\mathbf{s}(m)\mathbf{s}(m)^\dag\}\mathbf{A}^\dag+\E \{\mathbf{w}(m)\mathbf{w}(m)^\dag\}\nonumber\\
 &=\mathbf{A} \mathbf{R}_s\mathbf{A}^\dag+\mathbf{R}_n\, .
 \label{au}
 \end{align}
In the above equation, for uncorrelated sources, the signal correlation matrix, $\mathbf{R}_\mathrm{s}=\mathrm{diag}[\sigma_1^2,\ldots,\sigma_K^2]$, where $\sigma_k^2$ denotes the ${k}^{\underline{ \text{th}}}$ source power, and the noise 
 covariance matrix, $\mathbf{R}_n=\sigma^2 \mathbf{I}$. In practice, the auto-correlation matrix $\mathbf{R}$ will be replaced by the sample auto-correlation matrix averaged over $M$ time samples (or snapshots).  Since $K$ sources are considered, the largest $K$ eigenvalues of this auto-correlation matrix and their eigenvectors are related to the signal subspace. The other $N_\mathrm{tx}N_\mathrm{rx} -K  $ eigenvalues and eigenvectors correspond to the non-signal (noise) subspace.  In the development of the iterative DOA estimation algorithm, we assume a noise-free situation (i.e., $\sigma^2=0$). Later, the performance of the developed algorithm will be demonstrated in a noisy environment considering two different cases, where 1) there is prior knowledge of the noise variance, and 2) the noise variance of the system is totally unknown. 

Let us consider $N_\mathrm{tx}=N_\mathrm{rx}=K+1$. By using the non-signal subspace modeling in a uniform linear array \cite{santiago2013noise}, we can write, for $K$ sources, the non-signal subspace of the linear array along the axis representing the transmitter,
\begin{align}
    \mathbf{E}_{\mathrm{tx},K} & =\begin{bmatrix}
    \mathbf{E}_{\mathrm{tx},K-1}\\0
\end{bmatrix}-\exp(j\epsilon_{\mathrm{tx},K})\begin{bmatrix}
    0\\
    \mathbf{E}_{\mathrm{tx},K-1}
\end{bmatrix}  \nonumber\\
& =
    \mathbf{\tilde{E}}_{\mathrm{tx},K-1}
-\exp(j\epsilon_{\mathrm{tx},K})\mathbf{L}\mathbf{\tilde{E}}_{\mathrm{tx},K-1},
\label{nx}
\end{align} where 
$$ \mathbf{\tilde{E}}_{\mathrm{tx},K-1}=\begin{bmatrix}
    \mathbf{E}_{\mathrm{tx},K-1}\\
    0
\end{bmatrix}, \mathbf{L}= \begin{bmatrix}\mathbf{0}_{1\times k-1} & 1 \\\mathbf{I}_{K-1 \times K-1} & \mathbf{0}^T_{1 \times k-1} 
\end{bmatrix},$$ and $$\mathbf{E}_{\mathrm{tx},1}=\begin{bmatrix}
    1\\
    -\exp(j\epsilon_{\mathrm{tx},1})\end{bmatrix}.$$
The above equations show how one can recursively construct the non-signal subspace due to the $\ith{K}$ source from that of the $\ith{(K-1)}$ source. Similarly, the non-signal subspace of the linear array along the axis representing the receiver,
\begin{align}
    \mathbf{E}_{\mathrm{rx},K} & =\begin{bmatrix}
    \mathbf{E}_{\mathrm{rx},K-1}\\0
\end{bmatrix}-\exp(j\epsilon_{\mathrm{rx},K})\begin{bmatrix}
    0\\
    \mathbf{E}_{\mathrm{rx},K-1}
\end{bmatrix} \nonumber \\
& =
    \mathbf{\tilde{E}}_{\mathrm{rx},K-1}
-\exp(j\epsilon_{\mathrm{rx},K})\mathbf{L}\mathbf{\tilde{E}}_{\mathrm{rx},K-1},
\label{ny}
\end{align} where 
$$ \mathbf{\tilde{E}}_{\mathrm{rx},K-1}=\begin{bmatrix}
    \mathbf{E}_{\mathrm{rx},K-1}\\
    0
\end{bmatrix}, \quad \mathbf{E}_{\mathrm{rx},1}=\begin{bmatrix}
    1\\
    -\exp(j\epsilon_{\mathrm{rx},1})\end{bmatrix}.$$

Next, we prove a theorem to find a vector belonging to the non-signal subspace of the auto-correlation matrix $\mathbf{R}$ using (\ref{nx}) and (\ref{ny}).
\begin{theorem}
When there are $K$ sources and the number of antenna elements at the transmitter and receiver satisfies $N_\mathrm{tx}=N_\mathrm{rx}=K+1$, $$\mathbf{E}_{K}=\mathbf{E}_{\mathrm{tx}, K}\otimes \mathbf{E}_{\mathrm{rx}, K}$$ is a non-signal subspace vector of the auto-correlation matrix $\mathbf{R}=\E\{\mathbf{A}\mathbf{s}(m)\mathbf{s}^\dag(m) \mathbf{A}^\dag\}$.
\end{theorem}
\begin{IEEEproof}
Recall that vectors $\mathbf{E}_{\mathrm{tx},K}$ and $\mathbf{E}_{\mathrm{rx},K}$ are the non-signal subspace of auto-correlation matrix $\mathbf{R}_{\mathrm{tx}}=\E\{\mathbf{A}_\mathrm{tx}\mathbf{s}(m)\mathbf{s}^\dag(m) \mathbf{A}^\dag_\mathrm{tx}\}$, and  $\mathbf{R}_{\mathrm{rx}}=\E\{\mathbf{A}_\mathrm{rx}\mathbf{s}(m)\mathbf{s}^\dag(m) \mathbf{A}^\dag_\mathrm{rx}\}$, respectively. By definition, 
$$ \mathbf{A}_\mathrm{tx}^\dag\mathbf{E}_{\mathrm{tx},K}={\mathbf{0}_{K\times 1}},$$ and
$$ \mathbf{A}_\mathrm{rx}^\dag\mathbf{E}_{\mathrm{rx},K}={\mathbf{0}_{K\times 1}}.$$
The above equations yield,  
\begin{equation}
    \mathbf{a}_\mathrm{tx,k}^\dag\mathbf{E}_{\mathrm{tx},K}=\mathbf{a}_\mathrm{rx,k}^\dag\mathbf{E}_{\mathrm{rx},K}=0,
    \label{mm}
\end{equation}
where $k=1,\ldots,K.$ Note that $\mathbf{E}_{K}=\mathbf{E}_{\mathrm{tx},K}\otimes \mathbf{E}_{\mathrm{rx},K}$ will be the non-signal subspace vector of $\mathbf{R}$, if 
$$ \mathbf{A}^\dag\mathbf{E}_{K}={\mathbf{0}_{{K\times 1}}}.$$

First, we use PKP 1 to get
\begin{align*}
    \mathbf{A}^\dag\mathbf{E}_{K}    
    & = \begin{bmatrix}
    \mathbf{a}^\dag_\mathrm{tx,1} \otimes \mathbf{a}^\dag_\mathrm{rx,1}\\
    \vdots\\
    \mathbf{a}^\dag_\mathrm{tx,K} \otimes \mathbf{a}^\dag_\mathrm{rx,K}
\end{bmatrix} ( \mathbf{E}_{\mathrm{tx},K}\otimes \mathbf{E}_{\mathrm{rx},K}),
\end{align*}
and then we apply PKP 4 to write 

\begin{align*}
    \mathbf{A}^\dag\mathbf{E}_{K}    
    & = \begin{bmatrix}
    \mathbf{a}^\dag_\mathrm{tx,1} \mathbf{E}_{\mathrm{tx},K}\otimes \mathbf{a}^\dag_\mathrm{rx,1}\mathbf{E}_{\mathrm{tx},K}\\
    \vdots\\
    \mathbf{a}^\dag_\mathrm{tx,K} \mathbf{E}_{\mathrm{tx},K} \otimes \mathbf{a}^\dag_\mathrm{rx,K}\mathbf{\tilde{E}}_{\mathrm{rx},K}
\end{bmatrix}  \nonumber\\
    & = {\mathbf{0}_{K\times1}},
\end{align*}
where the equality is due to  (\ref{mm}) and we prove the theorem.
\end{IEEEproof}

In the next section, we will demonstrate how the non-signal subspace vector $ \mathbf{E}_{K}$ can be utilized to estimate DOAs of $K$ sources iteratively. This demonstration is performed using the following lemma on the property of $\mathbf{E}_{K}$ which is a straightforward application of PKP 3 with (\ref{nx}) and (\ref{ny}).

\begin{lemma}
Since the non-signal subspace vector $$\mathbf{E}_{K}=\mathbf{E}_{\mathrm{tx}, K}\otimes \mathbf{E}_{\mathrm{rx}, K}\,,$$ we have
 \begin{align}
     \mathbf{E}_{K}&=\mathbf{u}_0-\mathbf{u}_1\exp(j\epsilon_{\mathrm{rx},K})-\mathbf{u}_2\exp(j\epsilon_{\mathrm{tx},K})\nonumber\\
    & +\mathbf{u}_3\exp(j\epsilon_{\mathrm{tx},K})\exp(j\epsilon_{\mathrm{rx},K}),
     \label{noise}
 \end{align} where
 \begin{align}
 \label{u0}
 \mathbf{u}_0 & = \mathbf{\tilde{E}}_{\mathrm{tx},K-1}\otimes \mathbf{\tilde{E}}_{\mathrm{rx},K-1}\,,\\ \label{u1}
 \mathbf{u}_1 &= \mathbf{\tilde{E}}_{\mathrm{tx},K-1}\otimes (\mathbf{L} \mathbf{\tilde{E}}_{\mathrm{rx},K-1})\,,\\ \label{u2}
 \mathbf{u}_2 &= (\mathbf{L} \mathbf{\tilde{E}}_{\mathrm{tx},K-1})\otimes\mathbf{\tilde{E}}_{\mathrm{rx},K-1}\,,
\end{align}
 and
 \begin{align}
 \label{u3}
 \mathbf{u}_3 &= (\mathbf{L} \mathbf{\tilde{E}}_{\mathrm{tx},K-1})\otimes(\mathbf{L} \mathbf{\tilde{E}}_{\mathrm{rx},K-1})\,.
  \end{align}
\label{extralemma1}
\end{lemma}
Next, we prove a key property of the non-signal subspace vector $ \mathbf{E}_{K}$ that will be used later to develop the proposed iterative DOA estimation algorithm.

\begin{theorem}
When there are $K$ sources and the number of antenna elements at the transmitter and receiver satisfies $N_\mathrm{tx}=N_\mathrm{rx}=K+1$, $$\mathbf{u}_1-\exp(j\epsilon_{\mathrm{tx},K}) \mathbf{u}_3\,,$$ $$\mathbf{u}_2-\exp(j\epsilon_{\mathrm{rx},K}) \mathbf{u}_3\,,$$ and $$\mathbf{u}_0-\exp(j\epsilon_{\mathrm{tx},K}) \exp(j\epsilon_{\mathrm{rx},K}) \mathbf{u}_3$$ belong to the non-signal subspace.  \label{theorem2}
\end{theorem}

\begin{IEEEproof}
First, we will show that vector $\{\mathbf{u}_1-\exp(j\epsilon_{\mathrm{tx},K}) \mathbf{u}_3\}$ belongs to the non-signal subspace. In order to claim that, we need to prove
\begin{align}\label{d1}\mathbf{R}\{\mathbf{u}_1-\exp(j\epsilon_{\mathrm{tx},K}) \mathbf{u}_3\}={\mathbf{0}_{N_{\mathrm{tx}}N_{\mathrm{rx}}\times 1}}\end{align}

Applying PKP 1 and PKP 2 to (\ref{u1}) and (\ref{u3}), we write 
\begin{align*}
    & \mathbf{u}_1-\exp(j\epsilon_{\mathrm{tx},K})\mathbf{u}_3 \nonumber\\
    &=\{\mathbf{\tilde{E}}_{\mathrm{tx},K-1}-\exp(j\epsilon_{\mathrm{tx},K}) (\mathbf{L} \mathbf{\tilde{E}}_{\mathrm{tx},K-1})\}\otimes (\mathbf{L} \mathbf{\tilde{E}}_{\mathrm{rx},K-1}) \,,  
\end{align*}
 and then use (\ref{nx}) to get 
\begin{align*}
   \mathbf{u}_1-\exp(j\epsilon_{\mathrm{tx},K})\mathbf{u}_3 =\mathbf{E}_{\mathrm{tx},K}\otimes (\mathbf{L} \mathbf{\tilde{E}}_{\mathrm{rx},K-1})\,.
\end{align*}
In a noiseless scenario, substituting the value of the array auto-correlation matrix $\mathbf{R}$ from (\ref{au}), we write 
\begin{eqnarray*}
 \mathbf{R}\{\mathbf{u}_1-\exp(j\epsilon_{\mathrm{tx},K}) \mathbf{u}_3\} 
  = \mathbf{A} \mathbf{R}_s\mathbf{A}^\dag \{\mathbf{E}_{\mathrm{tx},K}\otimes (\mathbf{L} \mathbf{\tilde{E}}_{\mathrm{rx},K-1})\}\,. 
\end{eqnarray*}
Now, we sequentially apply PKP 1 and PKP 4  to obtain
\begin{align*}
    &\mathbf{A}^\dag\{\mathbf{E}_{\mathrm{tx},K}\otimes (\mathbf{L} \mathbf{\tilde{E}}_{\mathrm{rx},K-1})\}  \nonumber\\
    & = \begin{bmatrix}
    \mathbf{a}^\dag_\mathrm{tx,1} \otimes \mathbf{a}^\dag_\mathrm{rx,1}\\
    \vdots\\
    \mathbf{a}^\dag_\mathrm{tx,K} \otimes \mathbf{a}^\dag_\mathrm{rx,K}
\end{bmatrix} ( \mathbf{E}_{\mathrm{tx},K}\otimes \mathbf{L} \mathbf{\tilde{E}}_{\mathrm{rx},K-1}) \nonumber\\
    & = \begin{bmatrix}
    \mathbf{a}^\dag_\mathrm{tx,1} \mathbf{E}_{\mathrm{tx},K}\otimes \mathbf{a}^\dag_\mathrm{rx,1} \mathbf{L} \mathbf{\tilde{E}}_{\mathrm{rx},K-1}\\
    \vdots\\
    \mathbf{a}^\dag_\mathrm{tx,K}\mathbf{E}_{\mathrm{tx},K} \otimes \mathbf{a}^\dag_\mathrm{rx,K}\mathbf{L} \mathbf{\tilde{E}}_{\mathrm{rx},K-1}
\end{bmatrix}  \nonumber\\
    & = {\mathbf{0}_{K\times1}}.
\end{align*}
Note that the last equality is due to (\ref{mm}) and yields the desired result in (\ref{d1}).  Similarly, we can also prove that
 $$\mathbf{R}\{\mathbf{u}_2-\exp(j\epsilon_{\mathrm{rx},K}) \mathbf{u}_3\}={\mathbf{0}_{N_{\mathrm{tx}}N_{\mathrm{rx}}\times 1}}.$$
In a noiseless scenario, by the definition of non-signal subspace, we can write,
\begin{equation}
 \mathbf{R}  \mathbf{{E}}_{K}={\mathbf{0}_{N_{\mathrm{tx}}N_{\mathrm{rx}}\times 1}}.
 \label{cost}
\end{equation}
Inserting $\mathbf{E}_{K}$ from (\ref{noise}) into (\ref{cost}), we find
\begin{multline}
    \mathbf{R}[\mathbf{u}_0-\exp(j\epsilon_{\mathrm{rx},K})\mathbf{u}_1-\exp(j\epsilon_{\mathrm{tx},K})\mathbf{u}_2\\
     +\exp(j\epsilon_{\mathrm{tx},K})\exp(j\epsilon_{\mathrm{rx},K})\mathbf{u}_3]={\mathbf{0}_{N_{\mathrm{tx}}N_{\mathrm{rx}}\times 1}}.
\end{multline}
Since $\mathbf{u}_1-\exp(j\epsilon_{\mathrm{tx},K}) \mathbf{u}_3$ and $\mathbf{u}_2-\exp(j\epsilon_{\mathrm{rx},K}) \mathbf{u}_3$ belong the non-signal subspace, the above equation yields 
\begin{multline}
    \mathbf{R}\bracket{\mathbf{u}_0- \exp(j\epsilon_{\mathrm{tx},K})\exp(j\epsilon_{\mathrm{rx},K})\mathbf{u}_3}={\mathbf{0}_{N_{\mathrm{tx}}N_{\mathrm{rx}}\times 1}}\,,
\end{multline}
and therefore, vector $\mathbf{u}_0- \exp(j\epsilon_{\mathrm{tx},K})\exp(j\epsilon_{\mathrm{rx},K})\mathbf{u}_3$ also belongs to the non-signal subspace.
\end{IEEEproof}
The above theorem results in the following corollary which will guide us on how to iteratively estimate steering elements (or DOA angles) of the $\ith{K}$ user utilizing those of the other $K-1$ users along with the auto-correlation matrix $\bd{R}$.
\begin{corollary}
When there are $K$ sources in the scene and the number of antenna elements at the transmitter and receiver satisfies $N_\mathrm{tx}=N_\mathrm{rx}=K+1$, given the steering elements of $K-1$ sources (or $\{\bd{u}_i\}_{i=0}^{3}$), the steering elements of the $K^{\underline{\mathrm{th}}}$ source satisfy
\begin{equation}
    \exp(j\epsilon_{\mathrm{tx},K})=\frac{\mathbf{u}^\dag_1\mathbf{R}\mathbf{u}_1}{\mathbf{u}^\dag_1\mathbf{R}\mathbf{u}_3}\,,
\label{corollary1-eq1}
\end{equation}
\begin{equation}
    \exp(j\epsilon_{\mathrm{rx},K})=\frac{\mathbf{u}^\dag_2\mathbf{R}\mathbf{u}_2}{\mathbf{u}^\dag_2\mathbf{R}\mathbf{u}_3}\,,
\label{corollary1-eq2}
\end{equation}
and
\begin{equation}
 \exp(j\epsilon_{\mathrm{tx},K}) \exp(j\epsilon_{\mathrm{rx},K})=\frac{\mathbf{u}^\dag_0\mathbf{R}\mathbf{u}_0}{\mathbf{u}^\dag_0\mathbf{R}\mathbf{u}_3}\,. 
\label{corollary1-eq3}
\end{equation}
\label{corollary1}
\end{corollary}
Notice that in the above corollary (\ref{corollary1-eq1}) and (\ref{corollary1-eq2}) provide sufficient information regarding the steering elements of the $\ith{K}$ user. In the other words, (\ref{corollary1-eq3}) could viewed as redundant while (\ref{corollary1-eq1}) and (\ref{corollary1-eq2}) are available.    

\section{Iterative 2D DOA Estimation Algorithm (iDEA)}
In this section, using Corollary~\ref{corollary1}, we first develop the proposed iterative 2D DOA estimation algorithm (iDEA) assuming the number of antenna elements  $N_\mathrm{tx}=N_\mathrm{rx}=K+1$ in a noise-free environment. Next, we generalize iDEA for the number of antenna elements $N_\mathrm{tx}>K+1$ or  $N_\mathrm{rx}>K+1$. 

\subsection{When $N_\mathrm{tx}=N_\mathrm{rx}=K+1$} 
In one iteration, iDEA will execute the following steps.
\begin{itemize}
\item {\em Step 1:} Initialize  the $\ist{1}$ $(K-1)$ sources' steering elements and derive $\{\mathbf{u}_q\}^{q=3}_{q=0}$ assuming these $K-1$ sources form a set $\mathcal{Q}$.  
\item {\em Step 2:} Use (\ref{corollary1-eq1}) and (\ref{corollary1-eq2}) from Corollary~\ref{corollary1} to estimate the steering elements of the $K^{\underline{\mathrm{th}}}$ source. 
\item {\em Step 3:} 
\begin{itemize}
\item {\em a.} From set $\mathcal{Q}$, take out first $K-1$ users.
\item {\em b.} Label users 1 to $K-1$ as $2$ to $K$.
\item  {\em c.} Label the estimated $K^{\underline{\mathrm{th}}}$ source steering elements from Step 2 as user 1.
\item {\em d.} Update $\{\mathbf{u}_q\}^{q=3}_{q=0}$ from first $K-1$ users from set $\mathcal{Q}$
\item {\em e.} Go TO Step 2. 
\end{itemize}
\end{itemize}

When all the $K$ sources' steering elements in the scene are updated once, and we count that as one full iteration. We see that the complexity of iDEA will grow with each iteration. To control its complexity, we introduce a parameter, the maximum number of iterations $T$. The value of $T$ allows a trade-off between the complexity of the algorithm and the accuracy of the DOA estimation. If the iteration reaches $t=T$, iDEA immediately stops and calculates the DOAs for all the sources in the scene using the estimated steering elements. For the $\ith{k}$ source, the elevation angle can be estimated as

\begin{multline}  
\hat{\theta}_k =\sin^{-1} \paren{a_d[\{\angle \exp(j\hat{\epsilon}_{\mathrm{tx},k})\}^2+ \{\angle \exp(j\hat{\epsilon}_{\mathrm{rx},k})\}^2]^{\frac{1}{2}}},
\label{thest}
\end{multline}
 where $a_d= \lambda/(2\pi d)$  and $k=1,\ldots,K$. Using the above estimate of the elevation angle, the azimuthal angle of user $k$ is estimated as
\begin{multline}
    \hat{\phi}_k = \cos^{-1} \paren{\frac{a_d \angle \exp(j\hat{\epsilon}_{\mathrm{tx},k})}{\sin {\hat{\theta}_k}} } \bigcap \\ \sin^{-1}\paren{ \frac{a_d\angle \exp(j\epsilon_{\mathrm{rx},k})}{ \sin \hat{\theta}_{k}}}\,.
    \label{ph_est}
\end{multline}
Every steps of iDEA has been described with mathematical notations below:
\begin{algorithm}
\caption{Iterative 2D DOA Estimation Algorithm (iDEA) }\label{alg:cap}

\begin{algorithmic}[1]
\STATE Initialize $K-1$ sources' DOAs, $\theta_k$, and $\phi_k$ for $k=1,....,K-1$;
\STATE Calculate $\epsilon_{\mathrm{rx},k}$ and $\epsilon_{\mathrm{rx},k}$ for $k=1,....,K-1$, and form set $\mathcal{Q}$;
\STATE Derive $\{\mathbf{u}_q\}^{q=3}_{q=0}$ from these $K-1$ sources' steering elements;
\STATE Set the iteration index, $t=0$;

\WHILE{ $t < T$}

    \STATE $t=t+1$;
\FOR{$k=1,\ldots,K-1$}
    
    \STATE Find $K^{\underline{\mathrm{th}}}$ source steering elements, $\exp(j\epsilon_{\mathrm{tx},K})$ and $\exp(j\epsilon_{\mathrm{rx},K})$ using (\ref{corollary1-eq1}) and (\ref{corollary1-eq2}) from Corollary~\ref{corollary1};
\STATE From set $\mathcal{Q}$, take out first $K-1$ users;
\STATE Label users 1 to $K-1$ as $2$ to $K$;
\STATE Label the estimated $K^{\underline{\mathrm{th}}}$ source steering elements from Step 8 as user 1;
\STATE Update $\{\mathbf{u}_q\}^{q=3}_{q=0}$ from first $K-1$ users from set $\mathcal{Q}$;
\ENDFOR
\STATE Calculate  $\exp(j\epsilon_{\mathrm{tx},K})$, $\exp(j\epsilon_{\mathrm{rx},K})$ using Corollary~\ref{corollary1};
\ENDWHILE

\STATE Using the estimated steering elements, estimate all the DOAs, $\theta_k$, and $\phi_k$ for $k=1,....,K,$ utilizing (\ref{thest})- (\ref{ph_est});
\end{algorithmic}
\label{iDEA-algorithm}
\end{algorithm}

\subsection{When $N_\mathrm{tx}> K+1$ \, or \, $N_\mathrm{rx} > K+1$}
So far, we have assumed the number of elements at the transmitter and receiver, $N_\mathrm{tx}=N_\mathrm{rx}=K+1$. However, it is possible to have $N_\mathrm{tx}> K+1$ \, or \, $N_\mathrm{rx} > K+1$ in the scene. In that case, we divide the URA into different uniform sub-arrays, which have been extensively studied to enhance the DOA estimation precision and resolution \cite{wu2019doa,shi2017computationally,cantrell1981maximum}. These uniform sub-arrays should be of size $(K+1) \times (K+1)$, where the number of elements in both Tx and Rx axes is $K+1$. In this way, the total number of sub-arrays is $N_{\mathrm{sa}}=(N_\mathrm{tx}-K)\times (N_\mathrm{rx}-K) $. Now, we construct an averaged sample covariance matrix from all these sub-arrays, which will be utilized to perform a covariance matrix smoothing prior to applying the iterative algorithm. This averaged covariance matrix can be defined as
\begin{equation}
    \hat{\mathbf{R}}_{\mathrm{sa}}=\frac{1}{N_{\mathrm{sa}}} \sum_{i=1}^{N_{\mathrm{sa}}}\hat{\mathbf{R}_i}.
    \label{subarr}
\end{equation}

\section{Proof of Convergence} 

In this section, our goal is to prove the convergence of iDEA.  Recall that iDEA operates by updating a user's (say user $K$'s) steering elements, $\exp(j\epsilon_{\mathrm{tx}, K,\tau})$ and $\exp(j\epsilon_{\mathrm{rx}, K,\tau})$, where $\tau$ is the update-index. At different updates, different users are assigned which is denoted by $K$. In order iDEA to converge, we want $$ \lim_{\tau\rightarrow\infty}\epsilon_{\mathrm{m}, K,\tau} = \epsilon_{\mathrm{m}, K}\,,$$
where notation $\mathrm{m}$ is the generalized notation that could be either $\mathrm{tx}$ or  $\mathrm{rx}$. We begin the proof of convergence by proving the following lemmas. In order to do so, we would like to introduce following four terms  
\begin{eqnarray}
a_{\mathrm{m},k,\tau}&=&\tilde{\bd{E}}_{\mathrm{m},K-1,\tau}^{\dag}\bd{R}_{\mathrm{m},k}\tilde{\bd{E}}_{\mathrm{m},K-1,\tau} \label{term1}\,,\\
b_{\mathrm{m},k,\tau}&=&\tilde{\bd{E}}_{\mathrm{m},K-1,\tau}^{\dag}\bd{L}^\dag\bd{R}_{\mathrm{m},k}\tilde{\bd{E}}_{\mathrm{m},K-1,\tau}\,,\label{term2}
\end{eqnarray}
\begin{eqnarray}
c_{\mathrm{m},k,\tau}&=&\tilde{\bd{E}}_{\mathrm{m},K-1,\tau}^{\dag}\bd{R}_{\mathrm{m},k}\bd{L}\tilde{\bd{E}}_{\mathrm{m},K-1,\tau}\label{term3}\,,
\end{eqnarray}
and
\begin{eqnarray}
d_{\mathrm{m},k,\tau}&=&\tilde{\bd{E}}_{\mathrm{m},K-1,\tau}^{\dag}\bd{L}^\dag\bd{R}_{\mathrm{m},k} \bd{L}\tilde{\bd{E}}_{\mathrm{m},K-1,\tau}\,,\label{term4}
\end{eqnarray}
where  $\bd{R}_{\mathrm{m},k}= \bd{a}_{\mathrm{m},k}\bd{a}_{\mathrm{m},k}^\dag$.
\begin{lemma}
When the sources are uncorrelated, we have
\begin{eqnarray}
\mathbf{u}^\dag_{1,\tau}\mathbf{R}\mathbf{u}_{1,\tau} &=&\sum_{k=1}^{K}\sigma^2_k\times a_{\mathrm{tx},k,\tau}\times d_{\mathrm{rx},k,\tau}\,,\\ \label{cmd2}
 \mathbf{u}^\dag_{1,\tau}\mathbf{R}\mathbf{u}_{3,\tau} &=& \sum_{k=1}^{K}\sigma^2_k \times c_{\mathrm{tx},k,\tau} \times a_{\mathrm{rx},k,\tau} \,,\\  \label{cmd3}
  \mathbf{u}^\dag_{2,\tau}\mathbf{R}\mathbf{u}_{3,\tau} &=& \sum_{k=1}^{K}\sigma^2_k \times  b_{\mathrm{rx},k,\tau}\times d_{\mathrm{tx},k,\tau}\,, \\  \label{cmd4}
  \mathbf{u}^\dag_{2,\tau}\mathbf{R}\mathbf{u}_{2,\tau}&=&\sum_{k=1}^{K}\sigma^2_k \times  d_{\mathrm{tx},k,\tau}\times a_{\mathrm{ax},k,\tau}\,.
  \end{eqnarray}
  \label{lemma2}
\end{lemma}
\begin{IEEEproof}
Using (\ref{u1}), we obtain
\begin{align}
\mathbf{u}^\dag_{1,\tau}\mathbf{R}\mathbf{u}_{1,\tau} & = \{\mathbf{\tilde{E}}_{\mathrm{tx},K-1,\tau}\otimes (\mathbf{L} \mathbf{\tilde{E}}_{\mathrm{rx},K-1,\tau})\}^\dag \mathbf{R} \nonumber\\ &\times \{\mathbf{\tilde{E}}_{\mathrm{tx},K-1,\tau}\otimes (\mathbf{L} \mathbf{\tilde{E}}_{\mathrm{rx},K-1,\tau})\}\,.
\end{align}
Since we are considering noiseless scenario with uncorrelated sources, $\bd{R}= \sum_{k=1}^K\sigma^2_k\bd{A}_k\bd{A}_k^\dag$. Thus the above equation becomes 
\begin{align}
\mathbf{u}^\dag_{1,\tau}\mathbf{R}\mathbf{u}_{1,\tau} & =\sum_{k=1}^{K} \sigma_k^2\{\mathbf{\tilde{E}}_{\mathrm{tx},K-1,\tau}\otimes (\mathbf{L} \mathbf{\tilde{E}}_{\mathrm{rx},K-1,\tau})\}^\dag \mathbf{A}_k \nonumber\\ &\times  \mathbf{A}_k^\dag\{\mathbf{\tilde{E}}_{\mathrm{tx},K-1,\tau}\otimes (\mathbf{L} \mathbf{\tilde{E}}_{\mathrm{rx},K-1},\tau)\}\,.
\end{align}

Recall that $\mathbf{A}_k=\mathbf{a}_{\mathrm{tx},k}\otimes \mathbf{a}_{\mathrm{rx},k}$. Thus, PKP-1 and PKP-4 jointly allow us to write 
$$\{\mathbf{\tilde{E}}_{\mathrm{tx},K-1,\tau}\otimes (\mathbf{L} \mathbf{\tilde{E}}_{\mathrm{rx},K-1,\tau})\}^\dag \mathbf{A}_k$$ equals to 
$$\mathbf{\tilde{E}}_{\mathrm{tx},K-1,\tau}^\dag\mathbf{a}_{\mathrm{tx},k}\otimes  \mathbf{\tilde{E}}_{\mathrm{rx},K-1,\tau}^\dag\mathbf{L}^\dag\mathbf{a}_{\mathrm{rx},k} $$
and $$\mathbf{A}_k^\dag\{\mathbf{\tilde{E}}_{\mathrm{tx},K-1,\tau}\otimes (\mathbf{L} \mathbf{\tilde{E}}_{\mathrm{rx},K-1,\tau})\}$$ equals to 
$$ \mathbf{a}_{\mathrm{tx},k}^\dag\mathbf{\tilde{E}}_{\mathrm{tx},K-1,\tau}\otimes\mathbf{a}_{\mathrm{rx},k}^\dag\mathbf{L} \mathbf{\tilde{E}}_{\mathrm{rx},K-1,\tau} \,.$$
Therefore, after applying PKP-4 we rewrite 
\begin{align}
&\mathbf{u}^\dag_{1,\tau}\mathbf{R}\mathbf{u}_{1,\tau}  \\ \nonumber
& =\sum_{k=1}^{K} \sigma_k^2 \mathbf{\tilde{E}}_{\mathrm{tx},K-1,\tau}^\dag\mathbf{R}_{\mathrm{tx},k}\mathbf{\tilde{E}}_{\mathrm{tx},K-1,\tau}\\ \nonumber
& \qquad \qquad \otimes \mathbf{\tilde{E}}_{\mathrm{rx},K-1,\tau}^\dag\mathbf{L}^\dag\mathbf{R}_{\mathrm{rx},k}\mathbf{L} \mathbf{\tilde{E}}_{\mathrm{rx},K-1,\tau} \\ \nonumber 
 &= \sum_{k=1}^{K} \sigma_k^2 \times a_{\mathrm{tx},k,\tau} \otimes d_{\mathrm{rx},k,\tau}  \\
 &=\sum_{k=1}^{K} \sigma_k^2 \times a_{\mathrm{tx},k,\tau} \times d_{\mathrm{rx},k,\tau}\,. 
\end{align}
The last equality is due to the fact that $\otimes$ operation of two scalars is equivalent to their multiplication. The equality before the last one is due to (\ref{term1}) and (\ref{term4}). Similarly, we can derive (\ref{cmd2})-(\ref{cmd4}).
\end{IEEEproof}

Following the proof of Theorem 1 of \cite{santiago2013noise}, we get
$$\tilde{\bd{E}}_{\mathrm{m},K-1,\tau}^{\dag}\bd{a}_{\mathrm{m},k}= \Pi_{\iota=1}^{K-1}\paren{1-\exp\left\{j\paren{\epsilon_{\mathrm{m},\iota,\tau}-\epsilon_{\mathrm{m},k}}\right\}}\,,$$
and
\begin{align*}
    \tilde{\bd{E}}_{\mathrm{m},K-1,\tau}^{\dag}\bd{L}^\dag\bd{a}_{\mathrm{m},k}&= \exp\paren{j\epsilon_{\mathrm{m},k}}\times\\
    &\Pi_{\iota=1}^{K-1}\paren{1-\exp\left\{j\paren{\epsilon_{\mathrm{m},\iota,\tau}-\epsilon_{\mathrm{m},k}}\right\}}\,,
\end{align*}
which we apply to (\ref{term1}) - (\ref{term4}) for obtaining the following lemma.

\begin{lemma}
When the sources are uncorrelated, (\ref{term1}), (\ref{term2}), and (\ref{term3}) can be expressed in terms of the following trigonometric functions:
\begin{eqnarray}
a_{\mathrm{m},k,\tau}&=&4^{K-1}\Pi_{\iota=1}^{{K-1}}\sin^2\left(\frac{\epsilon_{\mathrm{m},k}-\hat{\epsilon}_{\mathrm{m},\iota,\tau}}{2}\right)\,,\label{term1a}\\
b_{\mathrm{m},k,\tau}&=&4^{K-1} e^{j{\epsilon}_{\mathrm{m},k}}\Pi_{\iota=1}^{K-1}\sin^2\left(\frac{\epsilon_{\mathrm{m},k}-\hat{\epsilon}_{\mathrm{m},\iota,\tau}}{2}\right)\nonumber\\
                     &=&e^{j{\epsilon}_{\mathrm{m},k}}a_{\mathrm{m},k,\tau}\,, \label{term2a}\\
c_{\mathrm{m},k,\tau}&=&4^{K-1} e^{-j{\epsilon}_{\mathrm{m},k}}\Pi_{\iota=1}^{K-1}\sin^2\left(\frac{\epsilon_{\mathrm{m},k}-\hat{\epsilon}_{\mathrm{m},\iota,\tau}}{2}\right)\nonumber\\
     &=&e^{-j{\epsilon}_{\mathrm{m},k}}a_{\mathrm{m},k,\tau}\,,\label{term3a} 
\end{eqnarray}
and $d_{\mathrm{m},k,\tau}=a_{\mathrm{m},k,\tau}\label{term4a}$.
  \label{lemma3}
\end{lemma}

Now we are almost ready to prove the convergence of iDEA. Before doing so, we need the following two cost functions for updating $\exp(j\epsilon_{\mathrm{tx}, K,\tau})$ 
\begin{eqnarray}\nonumber
    Q_{\mathrm{tx},K,\tau}&=&\{\mathbf{u}_{1,\tau}-\exp(j\epsilon_{\mathrm{tx},K,\tau}) \mathbf{u}_{3,\tau}\}^\dag\mathbf{R} \\     &\times&\{\mathbf{u}_{1,\tau}-\exp(j\epsilon_{\mathrm{tx},K,\tau}) \mathbf{u}_{3,\tau}\}\,,\label{cx}
\end{eqnarray}
and 
\begin{eqnarray}\nonumber
    {Q}_{\mathrm{tx},K,\tau+1}&=&\{\mathbf{u}_{1,\tau}-\exp(j\epsilon_{\mathrm{tx},K,\tau+1}) \mathbf{u}_{3,\tau}\}^\dag\mathbf{R} \\     &\times&\{\mathbf{u}_{1,\tau}-\exp(j\epsilon_{\mathrm{tx},K,\tau+1}) \mathbf{u}_{3,\tau}\}\,.\label{cx1}
\end{eqnarray}
Similarly, for updating $\exp(j\epsilon_{\mathrm{rx}, K,\tau})$ we define
\begin{eqnarray}\nonumber
    Q_{\mathrm{rx},K,\tau}&=&\{\mathbf{u}_{2,\tau}-\exp(j\epsilon_{\mathrm{rx},K,\tau}) \mathbf{u}_{3,\tau}\}^\dag\mathbf{R} \\     &\times&\{\mathbf{u}_{2,\tau}-\exp(j\epsilon_{\mathrm{rx},K,\tau}) \mathbf{u}_{3,\tau}\}\,,\label{cy1}
\end{eqnarray}
and
\begin{eqnarray}\nonumber
    {Q}_{\mathrm{rx},K,\tau+1}&=&\{\mathbf{u}_{2,\tau}-\exp(j\epsilon_{\mathrm{rx},K,\tau+1}) \mathbf{u}_{3,\tau}\}^\dag\mathbf{R} \\     &\times&\{\mathbf{u}_{2,\tau}-\exp(j\epsilon_{\mathrm{rx},K,\tau+1}) \mathbf{u}_{3,\tau}\}\,,\label{cy}
\end{eqnarray}

\begin{theorem}
Since $$Q_{\mathrm{m},K,\tau+1} <  {Q}_{\mathrm{m},K,\tau} \longrightarrow \lim_{\tau\rightarrow\infty} Q_{\mathrm{m},K,\tau} = 0\,,$$  for both m = tx and m = rx. Therefore, $$ \lim_{\tau\rightarrow\infty}\epsilon_{\mathrm{m}, K,\tau} = \epsilon_{\mathrm{m}, K}\,,$$ which ensures the convergence of iDEA.
\label{theorem3}
\end{theorem}
\begin{IEEEproof}
First, we demonstrate the desired result for m = tx by proving that
$${Q}_{\mathrm{tx},K,\tau} - Q_{\mathrm{tx},K,\tau+1} >0\,.$$
After some simplifications, (\ref{cx}) and (\ref{cx1}) jointly yield
\begin{multline} 
{Q}_{\mathrm{tx},K,\tau} - Q_{\mathrm{tx},K,\tau+1}\\
 = - \exp(-\epsilon_{\mathrm{tx},K,\tau+1})\mathbf{u^\dag}_{3,\tau}\mathbf{R}\mathbf{u}_{1,\tau} - \exp(\epsilon_{\mathrm{tx},K,\tau+1})\mathbf{u}_{1,\tau}^\dag\mathbf{R}\mathbf{u}_{3,\tau} \\
 +\exp(-\epsilon_{\mathrm{tx},K,\tau})\mathbf{u}_{3,\tau}^\dag\mathbf{R}\mathbf{u}_{1,\tau} + \exp(\epsilon_{\mathrm{x},K,\tau})\mathbf{u}_{1,\tau}^\dag\mathbf{R}\mathbf{u}_{3,\tau}.
    \label{con}
\end{multline}
According to  Corollary~\ref{corollary1}, $$\exp(\epsilon_{\mathrm{tx},K,\tau+1}) =  \frac{\mathbf{u}^\dag_{1,\tau}\mathbf{R}\mathbf{u}_{1,\tau}}{\mathbf{u}^\dag_{1,\tau}\mathbf{R}\mathbf{u}_{3,\tau}},$$ which yields the first line of the right-hand side of (\ref{con}) as
\begin{align}\nonumber
&- \exp(-\epsilon_{\mathrm{tx},K,\tau+1})\mathbf{u^\dag}_{3,\tau}\mathbf{R}\mathbf{u}_{1,\tau+1} - \exp(\epsilon_{\mathrm{tx},K,\tau})\mathbf{u}_{1,\tau}^\dag\mathbf{R}\mathbf{u}_{3,\tau} \\ 
&= -2\mathbf{u}^\dag_{1,\tau}\mathbf{R}\mathbf{u}_{1,\tau} = 2\sum_{k=1}^{K}\sigma^2_k a_{\mathrm{tx},k,\tau} a_{\mathrm{rx},k,\tau}\label{left-con}
\end{align}
Last equality is due to Lemmas~\ref{lemma2} and ~\ref{lemma3}. These two lemmas also yield the second line of the right-hand side of (\ref{con}) as
\begin{align}
& - \exp(-\epsilon_{\mathrm{tx},K,\tau})\mathbf{u}_{3,\tau}^\dag\mathbf{R}\mathbf{u}_{1,\tau} - \exp(\epsilon_{\mathrm{tx},K,\tau})\mathbf{u}_{1,\tau}^\dag\mathbf{R}\mathbf{u}_{3,\tau} \\ \nonumber
&=-\sum_{k=1}^{K} \sigma^2_k a_{\mathrm{tx},k,\tau} a_{\mathrm{rx},k,\tau}  (\exp(\epsilon_{\mathrm{tx},k}-\epsilon_{\mathrm{tx},K,\tau})\\ 
 &\qquad \qquad+ \exp(-\epsilon_{\mathrm{tx},k}+\epsilon_{\mathrm{tx},K,\tau})) \,,
\end{align}
which can be further simplified to write
\begin{align}\nonumber
& - \exp(-\epsilon_{\mathrm{tx},K,\tau})\mathbf{u}_{3,\tau}^\dag\mathbf{R}\mathbf{u}_{1,\tau} - \exp(\epsilon_{\mathrm{tx},K,\tau})\mathbf{u}_{1,\tau}^\dag\mathbf{R}\mathbf{u}_{3,\tau} \\ \label{right-con}
&=-2\sum_{k=1}^{K} \sigma^2_k a_{\mathrm{tx},k,\tau} a_{\mathrm{rx},k,\tau}\cos\paren{\epsilon_{\mathrm{tx},k}-\epsilon_{\mathrm{tx},K,\tau}}
\end{align}

Using (\ref{left-con}) and (\ref{right-con}) with  (\ref{con}), we get
\begin{multline} 
 {Q}_{\mathrm{tx},K,\tau} - Q_{\mathrm{tx},K,\tau+1}\\= 2 \sum_{k=1}^{K} \sigma^2_k a_{\mathrm{tx},k,\tau} a_{\mathrm{rx},k,\tau}  \paren{1-\cos\paren{\epsilon_{\mathrm{tx},k}-\epsilon_{\mathrm{tx},K,\tau}}}\\
= 4^{2K-1}\sum_{k=1}^{K} \sigma^2_k \sin^2\left(\frac{\epsilon_{\mathrm{tx},k}-\epsilon_{\mathrm{tx},K,\tau}}{2}\right)\times \\
\Pi_{m=1}^{{K-1}}\sin^2\left(\frac{\epsilon_{\mathrm{tx},k}-\epsilon_{\mathrm{tx},m,\tau}}{2}\right)\times\\\Pi_{n=1}^{{K-1}}\sin^2\left(\frac{\epsilon_{\mathrm{rx},k}-\epsilon_{\mathrm{rx},n,\tau}}{2}\right)
  \label{Fin}
\end{multline} 
Every term from $k=1$ to $k=K-1$ in then summation of (\ref{Fin}) is non-negative, thus
\begin{multline} 
{Q}_{\mathrm{tx},K,\tau} - Q_{\mathrm{tx},K,\tau+1}\\
\geq \sigma^2_K \sin^2\left(\frac{\epsilon_{\mathrm{tx},K}-\epsilon_{\mathrm{tx},K,\tau}}{2}\right)\times \\
\Pi_{m=1}^{{K-1}}\sin^2\left(\frac{\epsilon_{\mathrm{tx},K}-\epsilon_{\mathrm{tx},m,\tau}}{2}\right)\times\\\Pi_{n=1}^{{K-1}}\sin^2\left(\frac{\epsilon_{\mathrm{rx},K}-\epsilon_{\mathrm{rx},n,\tau}}{2}\right)
  \label{Fin1}
\end{multline} 

In a wireless environment with random DOA, the lower bound in (\ref{Fin1}) is positive definite (i.e. strictly $>0$)  unless  $\epsilon_{\mathrm{tx},K}=\epsilon_{\mathrm{tx},K,\tau}$. Since the cost function is bounded from below by zero and has been found to be strictly monotonically decreasing, it is deduced that the algorithm will converge \cite{SteinERami}, i.e.
$$\lim_{\tau\rightarrow\infty}\epsilon_{\mathrm{tx}, K,\tau} = \epsilon_{\mathrm{tx}, K}\,.$$

Similarly, using the following positive definite upper-bound 
\begin{multline} 
 {Q}_{\mathrm{rx},K,\tau} - Q_{\mathrm{rx},K,\tau+1}\\
\geq  \sigma^2_K\sin^2\left(\frac{\epsilon_{\mathrm{rx},K}-\epsilon_{\mathrm{rx},K,\tau}}{2}\right)\times \\
\Pi_{m=1}^{{K-1}}\sin^2\left(\frac{\epsilon_{\mathrm{tx},K}-\epsilon_{\mathrm{tx},m,\tau}}{2}\right)\times\\\Pi_{n=1}^{{K-1}}\sin^2\left(\frac{\epsilon_{\mathrm{rx},K}-\epsilon_{\mathrm{rx},n,\tau}}{2}\right)>0\,,
  \label{Fin3}
\end{multline} 
we claim $$\lim_{\tau\rightarrow\infty}\epsilon_{\mathrm{rx}, K,\tau} = \epsilon_{\mathrm{rx}, K}\,.$$ 

\end{IEEEproof}


\section{Numerical Results}
In this section, our objectives are, 1) to demonstrate the convergence of the proposed algorithm, iDEA, 2) to provide numerical simulation results to exhibit the performance of iDEA, and 3) to perform a complexity analysis between iDEA, and the 2D MUSIC algorithm in a MIMO system. In all the numerical examples, iDEA is implemented without any prior knowledge of the noise power, $\sigma^2$. However, its performance can be further improved by estimating $\sigma^2$ \cite{wu2021maximum, fei2008new,villano2013snr}, and subtracting it from the auto-correlation matrix. Of course, this estimation will increase algorithm complexity. In addition, the implementation of the baseline algorithm, 2D MUSIC in a MIMO system also does not hinge on the knowledge of $\sigma^2$; thus, the above implementation of iDEA will make the comparison fair.

For the simulations, we consider $K=2$ non-coherent far-field targets in a MIMO system  with $N_\mathrm{tx}=3$ transmitters, and $N_\mathrm{rx}=3$ receivers, which are uniformly placed (spacing, $d=\lambda/2$) along the $\mathrm{x}-$ , and the $\mathrm{y}-$ axis, respectively. Therefore, the angle between Tx and Rx axes, $\phi_{\mathrm{trx}}=90^{\circ}.$  Targets' DOA angels $(\theta_1,\phi_1)=(30^\circ,25^\circ)$, and $(\theta_2,\phi_2)=(70^\circ,80^\circ)$. Lastly, in order to run iDEA, we consider the initial values of the DOAs, $(\theta_{1,\mathrm{in}},\phi_{1,\mathrm{in}})=(10^\circ,10^\circ)$. For the 2D MUSIC algorithm in the MIMO system, we have used the search with $0.033^\circ$ angle precision for the DOAs. The above parameters are kept the same throughout the entire numerical study unless otherwise specified.

\subsection{Convergence as a Function of the Iterations}
\begin{figure}[t!]
\centering
    \includegraphics[width=\columnwidth]{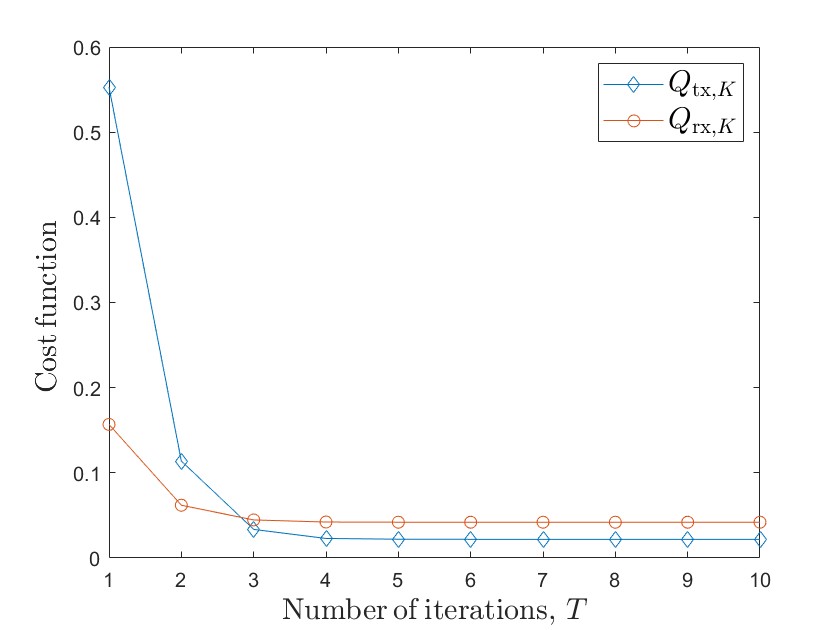}
    \caption{Convergence of iDEA for $M=10$.}
    \label{convg10}
\end{figure}

\begin{figure}[t!]
\centering
    \includegraphics[width=\columnwidth]{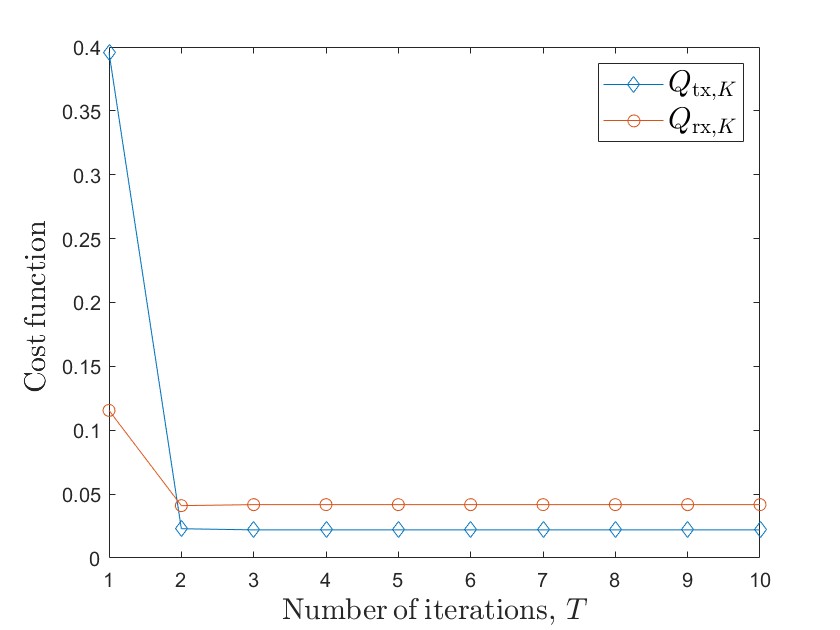}
    \caption{Convergence of iDEA for $M=500$.}
    \label{convg500}
\end{figure}

iDEA is an iterative algorithm, thus the demonstration of its convergence is important. In Figure \ref{convg10}, and \ref{convg500}, we plot the cost function $Q_{\mathrm{tx},K}$ in (\ref{cx}), and $Q_{\mathrm{rx},K}$ in (\ref{cy1}) as a function of the number of iterations ($T$) for two different number of samples, $M=10$, and $M=500$ at SNR = 30 dB.  As expected the convergence of iDEA depends on the number of samples, and a higher number of samples yields faster convergence. For example, when $M=10$, iDEA takes 4 iterations to get closer to the convergence value, and for $M=500$, this number is 2. Note that, after reaching the convergence, any increase in the number of iterations, $T$ may not contribute to the performance enhancement of the algorithm.

\subsection{Complexity Analysis}
Before we go any further evaluating the performance of our algorithm, we compare the
computational complexity of iDEA against 2D MUSIC for estimating DOAs in MIMO environments. We consider the complexity incurred after performing the matched filtering, and it is measured by the number of complex multiplications associated with the major operations. Note that one complex multiplication is equivalent to four real multiplications.

For iDEA, the cost of estimating the auto-correlation matrix is given by $(K + 1)^2M + M^2 + (K + 1)^2$ when $N_{\mathrm{tx}}=N_{\mathrm{rx}} = K + 1$  \cite{santiago2013noise}. if  $N > K + 1$, then $N_{\mathrm{sa}}$ sub arrays can be formed using (\ref{subarr}). Hence, the total calculation associated with the auto-correlation matrix is $\{(K + 1)^2M + M^2 + (K + 1)^2\}N_{\mathrm{sa}}$. Recall that iDEA operates by estimating the $\ith{K}$ source steering elements using (\ref{corollary1-eq1}) and (\ref{corollary1-eq2}) from Corollary~\ref{corollary1}. The cost of estimating these elements is $2KT\{(K+1)\times(K+2)+(K+1)+1\}$, where $K$ is the number of sources and $T$ is the number of iterations. Finally, after estimating all the steering elements, the cost associated with the estimation of the DOAs using (\ref{thest}), and (\ref{ph_est}) is $(7+4+4+1)/4=4$ complex multiplications \cite{cost,bc}. Therefore, the total cost associated with iDEA is $\{(K + 1)^2M + M^2 + (K + 1)^2\}N_{\mathrm{sa}}+2KT\{(K+1)\times(K+2)+(K+1)+1\}+4$. Notice that the complexity of the iDEA algorithm is primarily dictated by the number of samples ($M$), the number of iterations ($T$) and the number of sources ($K$).

Now, we focus on calculating the complexity of 2D MUSIC which performs eigenvalue decomposition (EVD) of the estimated auto-correlation matrix. EVD often is obtained from a singular value decomposition (SVD). Here, the cost associated with the estimation of the auto-correlation matrix is given by $N^2M + M^2 + N^2$, and the complexity associated with an SVD is $12N^3$, where $N=N_{\mathrm{tx}}N_{\mathrm{rx}}$\cite{santiago2013noise}. Then MUSIC estimates the DOAs by 2D angle search using the null space of the auto-correlation matrix. This cost is calculated as $K\times N_{\theta} N_{\phi}\{N(N-K)+1\}$, where $N_{\theta}$ and $N_{\phi}$ represent the searching point number on the azimuthal and elevation plane. Therefore, the total cost associated with the 2D MUSIC algorithm is $N^2M + M^2 + N^2+12N^3+K\times N_{\theta}N_{\phi}\{N(N-K)+1\}$. Notice that the virtual array size ($N$), the number of samples ($M$), the number of sources ($K$), and the product of the searching points ($N_{\theta}$ and $N_{\phi}$) dominate the complexity of the 2D MUSIC algorithm. The costs associated with implementing iDEA and 2D MUSIC are tabulated in Table \ref{t0} for comparison.

\begin{table}[t]
\begin{center}
\caption{Complexity analysis}
\vspace{-2pt}
\begin{tabular}{|c|c|}
\hline
{Algorithm} & Complexity (Number of complex multiplications)\\
\hline
\multirow{2}{*}{iDEA} & $\{(K + 1)^2(M+1) + M^2\}N_{\mathrm{sa}}$\\
 & $+2KT\{(K+1)(K+3)+1\}+4$\\
\hline
\multirow{2}{*}{2D MUSIC} & $N^2(M+1) + M^2 +12N^3$\\
 &$+KN_{\theta}N_{\phi}\{N(N-K)+1\}$\\
\hline
\end{tabular}
\label{t0}
\end{center}
\end{table} 

\begin{figure}[t!]
\centering
\includegraphics[width=\columnwidth]{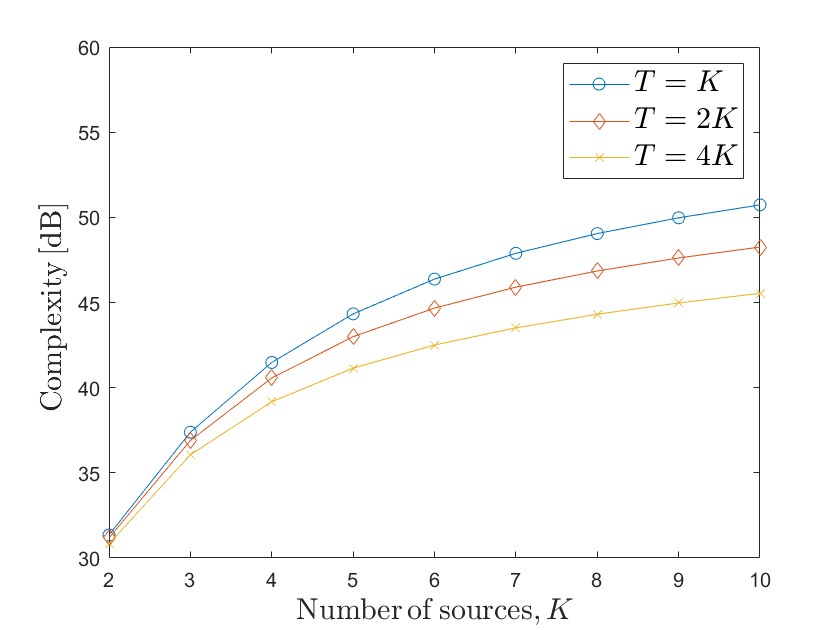}
\caption{Complexity of 2D MUSIC w.r.t. iDEA algorithm, where $N_{\mathrm{tx}}=N_{\mathrm{rx}}=K+1$, $M=50$.}
\label{complexity}
\end{figure}

Now, we use Table \ref{t0} to demonstrate the complexities of the DOA estimating algorithms by using a numerical example; see Figure \ref{complexity}, where the complexity of 2D MUSIC (w.r.t the iDEA algorithm) is plotted as a function of the total number of sources $K$, considering the number of Tx and Rx antennas, $N_{\mathrm{tx}}=N_{\mathrm{rx}}=K+1$, number of samples, $M=50$, and the number of iterations in iDEA for three different cases, $T=K$, $T=2K$, and $T=4K$. Here, the search in 2D MUSIC is conducted with $1^{\circ}$ precision of the DOA angles in both the elevation and the azimuthal planes. The results in  Figure \ref{complexity} suggest that depending on the system parameters our proposed iDEA is capable of offering significant complexity gain over 2D MUSIC. For instance, the offered gain is over $36$ dB for $K=3$, and $T=4K$.

\subsection{Performance as a function of SNR/the number of samples}
\begin{figure}[t!]
\centering
\includegraphics[width=\columnwidth]{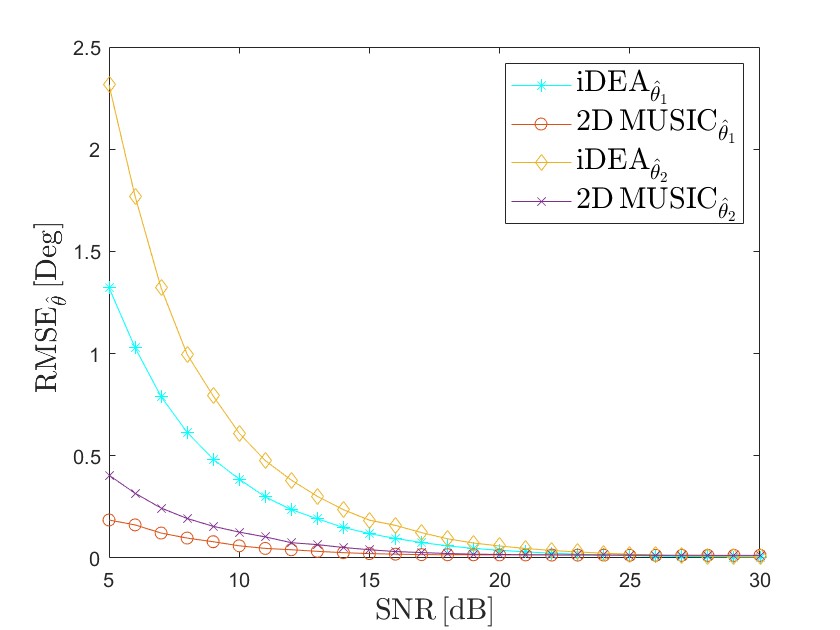}
    \caption{RMSE vs SNR for the elevation angle estimation.}
    \label{thsnr}
\end{figure}

\begin{figure}[t!]
\centering
    \includegraphics[width=\columnwidth]{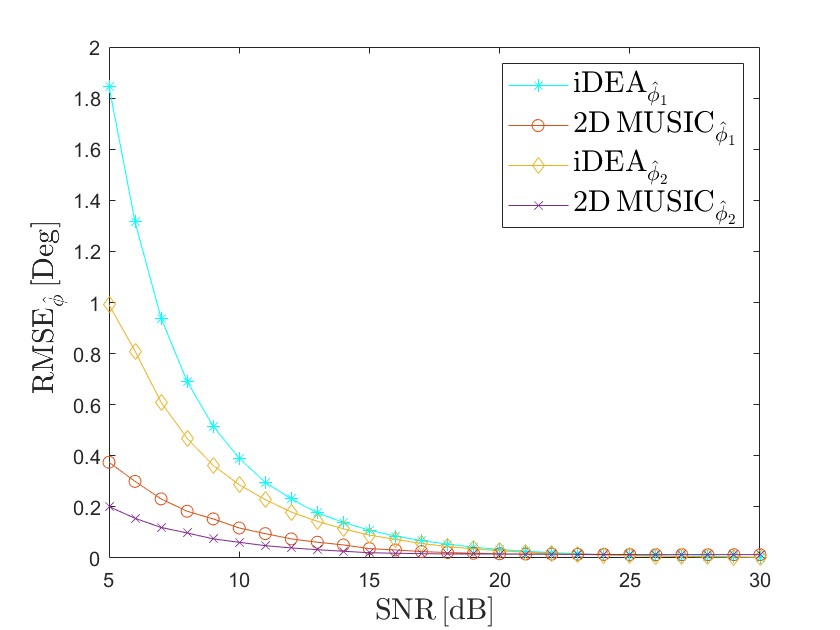}
    \caption{RMSE vs SNR for the azimuthal angle estimation.}
    \label{phsnr}
\end{figure}

\begin{figure}[t!]
\centering
    \includegraphics[width=\columnwidth]{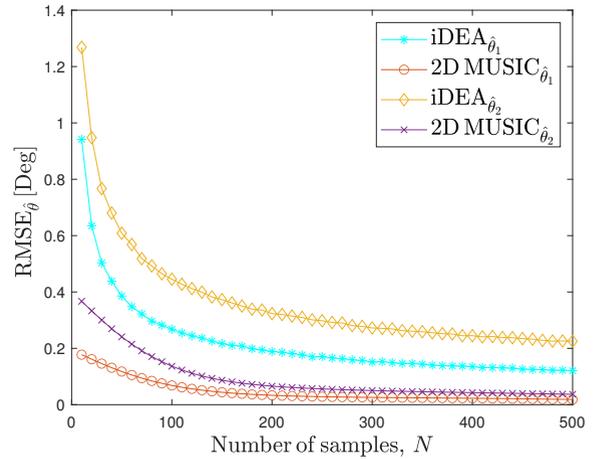}
    \caption{RMSE vs $N$  for the elevation angle estimation.}
    \label{N_th}
\end{figure}

\begin{figure}[t!]
\centering
    \includegraphics[width=\columnwidth]{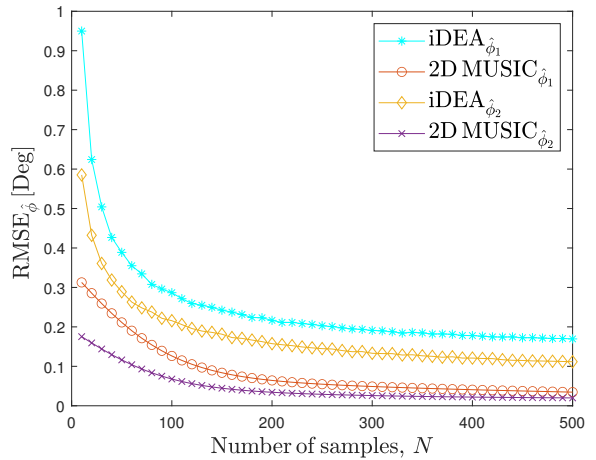}
    \caption{RMSE vs $N$ for the azimuthal angle estimation.}
    \label{N_ph}
\end{figure}

We now shift our focus to the performance iDEA. First, we demonstrate the root mean square error (RMSE) of iDEA and 2D MUSIC as a function of SNR for the number of iteration $T=6$ and the number of samples $M=50$; see Figure \ref{thsnr} and \ref{phsnr} for the elevation and azimuthal planes, respectively. Here, it can be noticed that, at low SNRs, the 2D MUSIC algorithm performs slightly better than iDEA. However, as SNR increases, the performance difference between the algorithms starts to diminish. As expected, similar observations are made after comparing the performance of the above two algorithms by varying the number of samples for $M$ for given SNR = 10 dB; see Figure \ref{N_th}, and \ref{N_ph}. Note that although iDEA does not have any prior knowledge of noise power in the environment, it performs well even in the presence of noise.

\begin{figure}[t!]
\centering
    \includegraphics[width=\columnwidth]{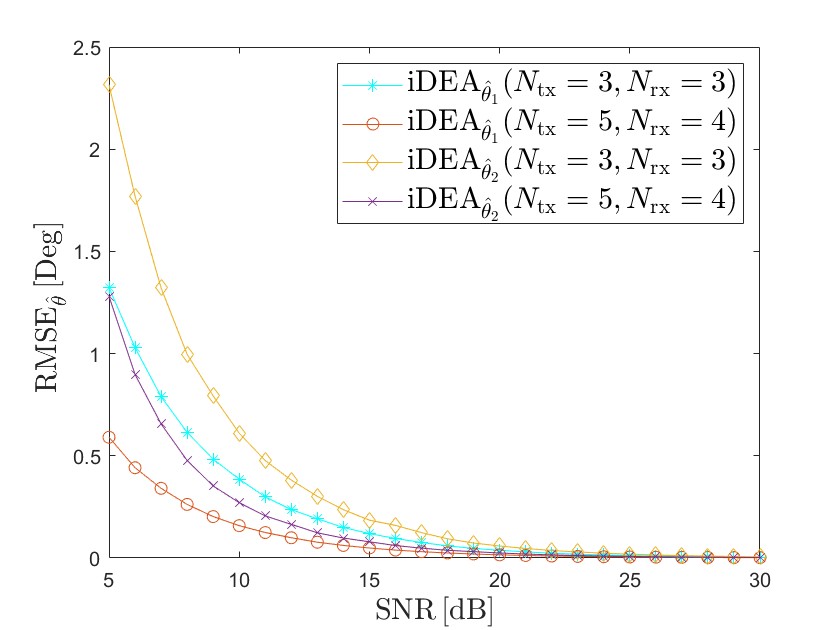}
    \caption{RMSE vs SNR for the elevation angle in two different virtual array scenarios.}
    \label{Na_th}
\end{figure} 

\begin{figure}[t!]
\centering
    \includegraphics[width=\columnwidth]{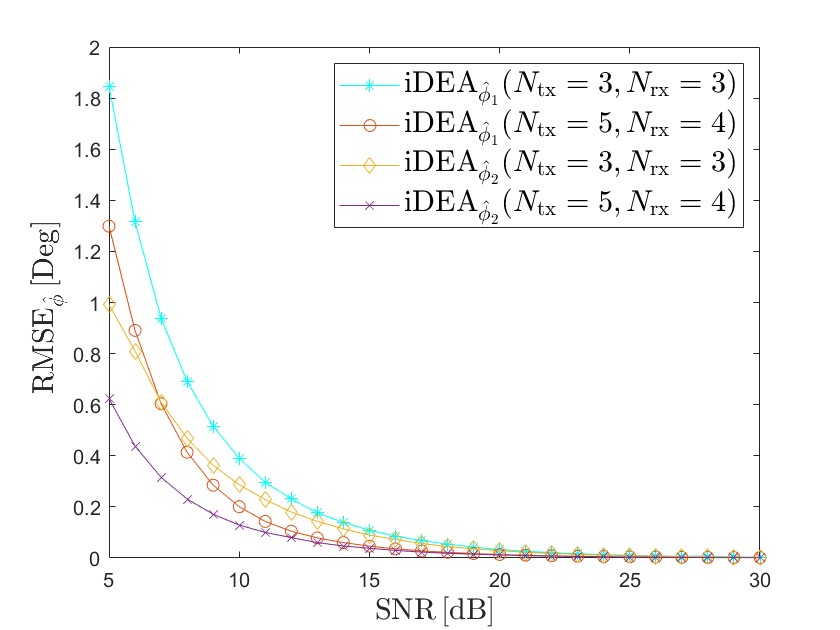}
    \caption{RMSE vs SNR for the azimuthal angle in two different virtual array scenarios.}
    \label{Na_ph}
\end{figure} 

\subsection{Performance as a function of the number of elements} 
 Finally, we investigate the performance of iDEA by varying the number of array elements. Here the same two source scenario ($K=2$) are considered for two different numbers of elements, $\{N_\mathrm{tx}=3, N_\mathrm{tx}=3\}$ , and $\{N_\mathrm{tx}=5, N_\mathrm{tx}=4\}$. Recall that when $N_\mathrm{tx}> K+1$  or  $N_\mathrm{rx} > K+1$, we form $N_\mathrm{sa}$ sub-arrays and average all $N_\mathrm{sa}$ auto-correlation matrices according to (\ref{subarr}) for obtaining the auto-correlation matrix for iDEA to operate. Therefore, for the first case, we obtain, the number of sub-arrays $N_\mathrm{sa}=1$, and for the second case $N_\mathrm{sa}=6$. Figure \ref{Na_th}, and \ref{Na_ph} depict RMSE for those 2 cases on both the elevation and the azimuthal planes while varying the SNR from 5 to 30 dB.  Here, we have used the number of iterations, $T=6$, and the total number of samples, $M=50$. It can be seen that the use of sub-arrays can significantly reduce the RMSE of the proposed algorithm iDEA. For example, at SNR = 5 dB, these reductions in RMSE are $55.33\%,44.70\%,29.65\%,37.09\%$ while estimating $\theta_1,\,\theta_2, \, \phi_1,\,\phi_2$, respectively. Note that the maximum number of sources iDEA can handle is $\min{\{N_{\mathrm{tx}},N_{\mathrm{rx}}\}}-1$. Thus, when $K<\min{\{N_{\mathrm{tx}},N_{\mathrm{rx}}\}}-1$, the use of sub-arrays can be used to improve the accuracy of the DOA estimation. However, this performance gain will be achieved at the cost of hardware and software complexity.

\section{Conclusion}
In this paper, a low-complexity iterative 2D DOA estimation algorithm, namely iDEA, has been developed for a MIMO system. This algorithm exploits the properties of theoretically derived eigenvectors associated with the non-signal subspace in a MIMO system when the number of signal sources is less than the number of elements in both transmitters and receivers. The convergence of the algorithm has been established both mathematically and numerically. Numerical results showed that the performance of the proposed algorithm was comparable to that of the 2D MUSIC algorithm in a MIMO system for a moderate to high SNR while exhibiting a significantly low complexity. For instance, this
complexity gain of the algorithm was over 36 dB for a number of sources, $K = 3$, and a number of iterations, $T = 4K$. In all the numerical examples, iDEA was implemented without the knowledge of the noise power. Its performance can be improved if the prior knowledge of the noise power is used or the sub-arrays are formed by using additional antennas at the transmitter or receiver or both. Using only two extra elements in the transmitter, and one element in the receiver, in a two-user system the RMSE of iDEA was reduced by $55.33\%,44.70\%,29.65\%,37.09\%$ while estimating $\theta_1,\,\theta_2, \, \phi_1,\,\phi_2$, respectively at SNR = 5 dB.
\bibliographystyle{ieeetran}
\bibliography{Reference.bib}
\begin{IEEEbiography}[{\includegraphics[width=1.0in,height=1.6in,clip,keepaspectratio]{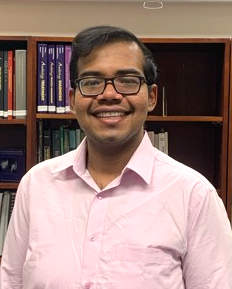}}]{Md Imrul Hasan} (Student Member, IEEE) received a B.Sc. degree from Bangladesh University of Engineering and Technology (BUET), in March 2016 with a major in Electrical and Electronic Engineering. He completed his master’s degree in May 2022 and now pursuing his doctoral degree with a major in Electrical Engineering from The University of Texas at Dallas. MD IMRUL HASAN joined LitePoint Corporation, a Teradyne company, in October 2022 as a Wireless Signal Processing Engineer, where he is currently an integral part of the engineering development team. Here, he develops and implements advanced signal-processing algorithms for wireless technologies.
\end{IEEEbiography}
\begin{IEEEbiography}[{\includegraphics[width=1.0in,height=1.2in]{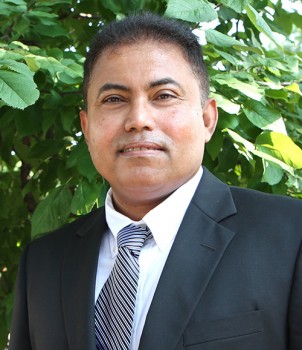}}]{Mohammad Saquib} (Senior Member, IEEE) received the B.Sc. degree from the Bangladesh University of Engineering and Technology, Bangladesh, in 1991, and the M.S. and Ph.D. degrees from Rutgers University, New Brunswick, NJ, USA, in 1995 and 1998, respectively, all in electrical engineering. He worked as a member of the technical staff with the Massachusetts Institute of Technology Lincoln Laboratory, and as an Assistant Professor with Louisiana State University, Baton Rouge, LA, USA. He is a Professor with Electrical Engineering Department, The University of Texas at Dallas, Richardson, TX, USA. His current research interests include the various aspects of wireless data transmission, radio resource management, and signal processing techniques for low-cost radar applications.
\end{IEEEbiography}

\EOD

\EOD

\end{document}